# The extended formulation for KZM

Author: Wei Wen[1*], Shanhua Zhu[1], Yi Xie[2,3], Wei Wu[2,3], Baoquan Ou[2,3], Pingxing Chen[2,3*]

2016.09

## Abstract

KZM is applied broadly in the cosmology and condensed matter field to analyze topological defects. However, the drawback, which is that KZM can only deal with the defects formed in linear and global quenching conditions, greatly limits its application field. It is almost blank in nonlinear or local quenching area for KZM[(3b)]. Even for the seemingly "had been solved" inhomogeneous structure, it is hard to say satisfactory. We propose a new extended formulation that can well predict the scale of topological defects in both nonlinear and local quenching conditions. We show it is not necessary to deal with different cases with different approximation methods, the defects varying with quenching rate has unified form, which is in better agreement with experimental and simulation results. Additionally, guiding with this theory, we find oscillation quenching methods that can effectively suppress the defects produced in ion trap. This work will provide a new way for us to find an effective method about very fast quantum state manipulations on adiabatic quantum computation besides the broad applications on continue phase transition areas.

Kibble-Zurek mechanism (KZM) is the theory to calculate the relationship between topological defects and linear driven rate under the continuous phase transformation with finite linear driving rate in the non-equilibrium dynamical system. In fact, it is one of few physics theories that can be applied in both macroscopic and quantum phase transition. The early thought of KZM is formed by T. W. B. Kibble from 1976s to 1980s in his two papers. In his papers, Kibble proposed a model to estimate the density of cosmic galaxy, cosmic string, cosmic domain, *etc*.[1,2]. By Kibble's thought, the cosmic domains will be in the

[1] *College of Science*, Hunan university of technology, Zhuzhou, China.
[*] To whom correspondence should be addressed, E-mail: chuxiangzi@semi.ac.cn
[2] *College of Science*, National University of Defense Technology, Changsha, China
[3] *Interdisciplinary Center of Quantum Information,* National University of Defense Technology, Changsha, China
[*] To whom correspondence should be addressed, E-mail: pxchen@nudt.edu.cn

"frozen area" when they very approach to critical points and in the "adiabatic evaluation" status when they are far away critical points. W.H. Zurek pointed the parallels between the vortex lines formed in super flux and cosmic string formed in earlier cosmos, and used the Kibble's thought to calculate the density of vortex lines in the superfluid of $^4$He firstly. For decades of development, the thought of Kibble dealing with the formation of cosmic string have been generalized to the condensed matter, such as the superfluid of $^4$He[3-5] and $^3$He[6,7], the spontaneous nucleation of magnetic flux[8] and Josephson Tunnel Junctions[9-13] , the kinks of the ion chain[14-17] and so on. These works support the thought of Kibble and Zurek, and form the basement of Kibble-Zurek Mechanism(KZM). The thought of KZM can be simply summed up three assumptions, 1). The evaluation assumption. The system will undergo two kinds of evolutions—the adiabatic evolution at neighboring region of critical points and the impulse evolution at the far away of critical points; 2). The boundary assumption. The boundary of adiabatic evolution and impulse evolution is determined by equation $\hat{t} = \tau(\hat{t})$, where the $\tau$ is the relaxation time; 3). The defects-estimation assumption. The density of defects can be written as $d \sim 1/\hat{\xi}$, where the $\xi$ is the correlation length. According to these simple assumptions, KZM shows us the defects exhibit the scaling behavior and it is directly proportional to the fractional power of the quenching rate $1/\tau_Q$[3,16-21] (see in *table*1) and verified in many experiments [4,8,14,15].

## KZM and Extended formulation

However, the progress of KZM is mainly made in the conditions of linear and global quenching. It is almost blank in the nonlinear or local quenching [3a,3b]. Even if in the linear and global quenching condition, the predicts of KZM are hard to say be satisfied. For example, 1). For the system with finite scale, such as ion chain in ion trap, we can observe the saturated quenching [4] as quenching rate increasing, but KZM doesn't give the calculation method; 2). For the inhomogeneous structure, the predictions of KZM are not in good agreement with actual results. In 2013, K. Pyka, Mehlstaubler, et al. used highly reproducible experiment to test the scaling of defects formed in the harmonic ion trap predicted by KZM[15], and supported the data shown in *fig.*1(b1). According the prediction of KZM, the density of kinks formed in the inhomogeneous ion chain is proportional to the $1/\tau_Q$ with the fractional power $1/3$, $4/3$ and 8/3 respectively as $\tau_Q$ increases under the underdamped condition[17]. But the data shown in *fig.*1(b1) is of the smooth curve $f: \tau_Q \rightarrow d$; 3). KZM uses lots of semi-quantitative

---

[3a] Unlike the global quenching, under which all parts of a system have the same quenching rate, the local quenching means the quenching rate acting on different parts of a system is varying.

[3a] As we found, for the nonlinear quenching with the form $\epsilon = |t/\tau_Q|^r \mathrm{sign}(t)$, KZM is also applicable (see supplementary material).

[4] Saturated quenching is the status that rate of defects producing keep almost still no matter how to increase the quenching rate.

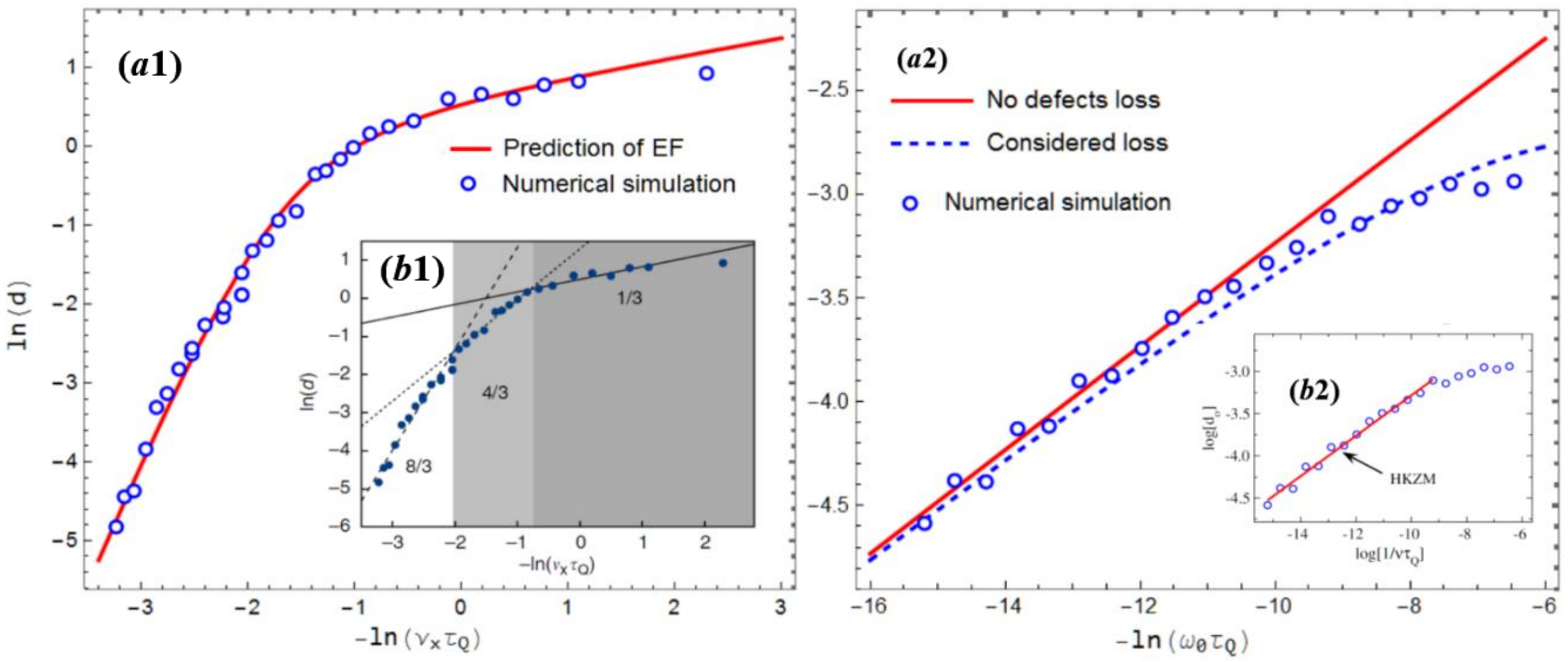

**Figure 1. The prediction of Extended formulation (EF) in the ion trap.** (a1) and (b1) are the results in the harmonic ion trap; (a2) and (b2) are the results in the nonharmonic ion trap, where the ion chain can be treated as the homogeneous structure. In the (a1), we give the density of defects calculated by the extended formulation for KZM. The parameters we chose are the same with the (b1), which is from the reference *Nat Commun* **4**, 2291(2013). In the (b1), the defect function is divided into three lines to meet the KZM result, but we show it is the smooth curve and quite good agreement with function depended by Eq. (1). The figure about the defects in the harmonic ion trap varying with the large range of the $\tau_Q$ is shown in *fig.*2(b). In (a2) we show the curve that calculated exactly from the extended formulation. For being compared with KZM, the parameters chosen in (a2) are the same with (b2), which supported by *New Jour. Phys.* **12** 115003(2010). This curve is with the formula $d \simeq k\tau_Q^{-1/4}$. If we consider the defects loss, then the $d \simeq f_p(k\tau_Q^{-1/4})k\tau_Q^{-1/4}$, where $f_p(x) = px/(\sum n(px)^n)$, the $p$ here $\approx 2.89$ caused by non-uniform distribution in actual nonharmonic ion trap. The figure about the defects in homogenous ion chain varying with the large range of the $\tau_Q$ is shown in *fig.*2(a).

parameters which make it be not the exact theory. On the interpretation for the actual results, like overdamped, underdamped are used in KZM, but it neither gives the strict point between them nor shows how defects vary with quenching rate apart from these two conditions. In KZM, the method to estimate defects rate in inhomogeneous structure is different from that in homogeneous structure, which can be seen in Table 1. The front velocity and characteristic velocity are introduced in inhomogeneous structure, and the semi-quantitation, much less, much more are also pull-in. Hence, in *fig.*1(b1), the KZM points out the power scalar but cannot give the exact boundary value where is with the fractional power $1/3$, where is with $4/3$ and 8/3.

Can the defect function of quenching rate have a unified form? Can we predict the defect rate using a method no matter whether in linear quenching or not, homogeneous structure or not and overdamping or not? In this article, we show a possible unified form, extended for-mulation, in *Table* 1, that can solve these problems. For example, per the extended formulation, we can see, it is not necessary to classify the different quenching conditions, the fractional power of the quenching rate is a continuous change of quenching rate. We show the actual results showing in *fig.*1(b1) can be well fitted by the following partial differential equation, Eq. (1). In the harmonic ion chain, the density of defects is $d(\tau_Q) = L/\xi(\tau_Q, v(\tau_Q))$, where $\xi(\tau_Q, v)$ is satisfied

$$\frac{\partial_v \xi}{2v\tau_Q} = -\frac{4v\gamma^2(v^2+\alpha\gamma^2)}{(\alpha\gamma^2-v^2)^3} + 3\chi\xi(L^2-\xi^2)^2\,\partial_v\xi. \tag{1}$$

where, $\gamma = \xi_1/\tau_1$, $\alpha = 4\delta_0/\eta^2$, $\chi = 189Q^2\zeta(3)N^3\tau_Q/512\pi m L^3\delta_0$.

In fact, the extended formulation supports a more power method to calculate the defects formed in continuous phase transition and get more conclusions. Taking the ion trap system as an example, per the extended formulation, we can get at least three conclusions: 1). For linear driving rate and global quenching in homogenous ion chain, the extended formulation is compatible with KZM, but contain more details than KZM; 2). For inhomogeneous ion chain, the extended formulation supports the better predicts than KZM; 3). For non-linear driving or local quenching, the extended formulation can be also applied, and it shows the oscillation quenching designed thought Extended formulation can greatly suppress the defects producing in the ion chain. These results are suitable for not only ion trap system, but also other systems or models, including LZ model, superfluid of $^4\mathrm{He}$ and $^3\mathrm{He}$, and so no.

**The basements of extended formulation**

The extended formulation is based on new three assumptions: 1). Boundary assumption, $\hat{\tau}_N = f(\tau_Q)\hat{t}$; 2). Propagation velocity assumption, $v_p = \xi_N/\tau_N$; 3). Defects-estimation assumption, $d = 1/\langle\hat{\xi}\rangle$, where $\hat{\xi} = \int_0^{\hat{\tau}_N} v_p \mathrm{d}t$ (see in *table* 1(a)).

To explain how these assumptions gotten in extended formulation in *table* 1, we give a brief explanation with stochastic force. According the idea of Kibble and Zurek, the state of system can readjust to the new set of thermodynamic parameters and the system can be regarded as evolving close to equilibrium, whereas the state will lag and the situation will be close to the instantaneous phase transition, then they choose the time point $\tau(\hat{t}) = \hat{t}$ as a suitable boundary point for these two distinct processes. However, neither the deep interpretation nor the fundamental is given for this. Therefore, it leaves a contradiction to interpret the phase transition under quick quenching. Under quick quenching, a system will be in the status analogous to supercooling shortly when the system is across the critical point and in the frozen area that KZM mentioned. The analogous supercooling status has the strong correlation[22,23] and it is seemed conflicts the limit correlation $\xi(\hat{t})$ that the KZM used. How to solve this confliction?

We find the equation $\tau(\hat{t}) = \hat{t}$ is the result of the competition between the order structural transition and the disorder stochastic breaking. In the analogous supercooling status, the system has strong correlations but fragile structure. With a slight disturbance caused by stochastic force, a nucleate sit of structural transition will be formed and the others particles around the nucleation will adjust their position one by one for strong correlations, and then the order structural transition is beginning with limit propagation velocity $v$. At the same time, the ubiquitous stochastic force $\varepsilon(t)$ in some areas of the system will produce the other nucleation sites and cause the disorder stochastic breaking if the impact of the first nucleation forming has no enough time to spread to these areas. Then the defects will be formed.

Stochastic forces $\varepsilon(t)$ satisfy the fluctuation-dissipation relation $\langle \varepsilon_i(t_1)\varepsilon_j(t_0)\rangle = 2m\eta k_b T\delta_{i,j}\delta(t_1 - t_0)$. Using $\tilde{\varepsilon}_i(\Delta t_j)$ to express the average stochastic force $\tilde{\varepsilon}_{i,j}$ under the $i$th particles in very small time interval $\Delta t_j$, then the probability about the $\varepsilon_i(\Delta t_j)$ can be written as the Gaussian function[24-26], $\mathrm{Prob}(\varepsilon_i(\Delta t_j)) = \exp -(\varepsilon_{i,j}^2 \Delta t/4k_b T\eta)$. Some stochastic force in time sequence will produce enough impulse to make the particle overcomes the potential barrier and cause the stochastic breaking if $\varepsilon_{i,j}\Delta t_j \geq 2m\omega^2 \tilde{l}\sqrt{\tau\Delta t_j}$, where $\tau$ is the relaxation time[17,27]. the probability of the stochastic breaking is $prob(\varepsilon') = \mathcal{N}\int_{\varepsilon'}^{+\infty} \varepsilon_{i,j}^2 \exp -(3\varepsilon_{i,j}^2 \Delta t/4k_b T\eta) d\varepsilon_{i,j}$, where $\mathcal{N}$ is normalization constant. Considering the long-time interval $\tau_N$, the average number that cause the stochastic breaking is $\langle n\rangle = (\tau_N/\Delta t_j) prob$. Let $\langle n\rangle = 1/N$, and then (see the supplementary material)

$$\tau_N \backsimeq \exp\left(\frac{m\omega_{t,c}^4}{8k_b T\mathcal{A}}\right)\frac{8k_b T\mathcal{A}\gamma\omega_{t,c}^{1/2}}{Nm\eta^3}\tau = c_N\tau. \tag{2}$$

The constant $c_N$ in Eq. (2) does not change the trend of defects varying with quenching rate in KZM, but it makes the time interval $\tau_N$ meaningful. The $\tau_N$ is the life time of particles in the supercooling system. It is the quantity to weigh the time that the parts of a system can be against the disorder. If the evolution time of a particle is beyond its lifetime $\tau_N$ and the effect of the first nucleation sit does not reach here, then the particle will form the disorder stochastic breaking with high probability. In the time interval $\tau_N$, the length that the effect of nucleation sit can propagate is renamed $\xi_N$. To keep the consistence with the KZM in homogeneous structure, the $\xi_N$ should be proportional to the correlation length $\xi$. Therefore, the propagation velocity of nucleation effect can be written as

$$v_p = \frac{\xi_N}{\tau_N} \propto \frac{\xi}{\tau}. \tag{3}$$

The propagation velocity of the nucleation effect is the state function and depends on the reduced parameter $\epsilon = (M - M_c)/M_c$, where $M$ can be temperature, frequency, magnetic field and so on. For the static status, the "boundary point" is $\tau_N(\hat{t}) = \hat{t}$, but for the driving steered status, the "boundary point" should be substituted by

$$\begin{gathered} \hat{\tau}_N = \frac{1}{\beta^{1/2}\dot{\epsilon}^{1/2} + 1}\hat{t}, \\ \beta = 4\left(\frac{a_0}{\xi_1}\right)^3 \frac{\eta}{\delta_0^{1/2}}\tau_1 \end{gathered} \tag{4}$$

where $a_0$ is the distance of particles in homogenous structure. Eq. (4) is compatible with KZM in homogenous continuous medium and slow quenching. In fact, for the continuous medium, such as the superfluid of $^4He$, $a_0 = 0$, and Eq. (4) transforms to the formula of KZM $\hat{\tau}_N = \hat{t}$. Additionally, under slow quenching rate, $\tau_Q \gg \beta$, and $\hat{\tau}_N = \hat{t}$ is also held.

Under the quenching, the propagation velocity change in every time, the distance that the effect of nucleate sites can propagate to should transform to integral formula

$$\hat{\xi} = \int_0^{\hat{\tau}_N} v_p \mathrm{d}t. \tag{5}$$

This equation is one of basements for extended formulation and makes the calculation for defects be not a simple algebra computation in inhomogeneous structure or nonlinear, local quenching conditions. But for the linear quenching and homogeneous structure, Eq. (5) will transform into the algebra computation and be compatible with the KZM (see the prove in the supplementary material).

**I). For linear driving rate and global quenching in homogeneous ion chain.**

In the ion chain, the expression of $\tau$ should be written more exact as Eq. (6) since it comes from the Euler–Lagrange equation[17,27],

$$\tau = \frac{\tau_0}{\sqrt{1 + \alpha|\epsilon|} - 1} \propto \tau_N, \tag{6}$$

Let $\tau_1 = c_N\tau_0 \sim c_N/\eta$, then we can see, if $\alpha|\epsilon| \gg 1$, namely $\delta_0|\hat{t}|/\tau_Q \gg \eta^2$, then, $\tau_N \approx \tau_1/|\epsilon|^{1/2}$. This is the underdamping condition that KZM mentioned; Conversely, if $\alpha/\tau_Q \ll 1$, namely $\delta_0|\hat{t}|/\tau_Q \gg \eta^2$, then, $\tau_N \approx \tau_1/|\epsilon|$. It is the overdamping condition that KZM speaking. That is to say Eq. (6) can be compatible with the KZM formulation in some extreme

**Table 1. The comparison of expressions between KZM and extended formulation in the ion trap.**

| (*a*) | KZM | Extended formulation |
|---|---|---|
| **In the homogeneous structure** | $d\sim L/\hat{\xi};\ \hat{\xi}=\xi(\hat{t});\ \hat{t}=\tau(\hat{t});$<br>$\xi=\begin{cases}\xi_0\lvert\epsilon\rvert^{-1/2}; in\ LT\\ \xi_0\lvert\epsilon\rvert^{-1/2}. in\ RG\end{cases}$<br>$\tau=\begin{cases}\tau_0\lvert\epsilon\rvert^{-1}; underdamping\\ \tau_0\lvert\epsilon\rvert^{-1/2}. overdamping\end{cases}$ | $d=L/\langle\hat{\xi}_{a_i}\rangle$ ;<br>$\hat{\xi}_{a_i}=\int_0^{\hat{\tau}}v_p(x_{a_i},t)\mathrm{d}t$ ;<br>$v_p=\xi_N(x_{a_i},t)/\tau_N(x_{a_i},t);$<br>$\hat{t}=f(\tau_Q)\tau_N(\hat{t});$<br>$f(\tau_Q)=\frac{\dot{\epsilon}^{1/2}}{(\beta\dot{\epsilon})^{1/2}+1};$ |
| **In the inhomogeneous structure** | $v_{\mathrm{F}}\sim\frac{\partial_t\delta(x,t)}{\partial_x\delta(x,t)}$<br>$\hat{v}_x=\hat{\xi}_x/\hat{\tau}_x=a\omega_0(\delta_0/\eta^3\tau_Q)^{1/2}$<br>$d=\begin{cases}d_u\sim\begin{cases}1/\hat{\xi};\quad if\ v_f\gg v_x\\ 2X/\hat{\xi};\ if\ v_f\sim v_x\\ (\tau_Q)^{-8/3};if\ v_f\ll v_x\end{cases}\\ d_o\sim 2X_o/\hat{\xi}_o\propto(\tau_Q)^{-1}\end{cases}$ | $\tau_N=C\tau$<br>$=\tau_1\left(\sqrt{1+\alpha\lvert\epsilon\rvert}-1\right)^{-1}$<br>$\xi_N\propto\xi=\begin{cases}\xi_0\lvert\epsilon\rvert^{-1/2}; in\ LT\\ \xi_0\lvert\epsilon\rvert^{-1/2}. in\ RG\end{cases}$<br>($a_i$ is the inter − particle distance) |

| (*b*) | KZM | Extended formulation |
|---|---|---|
| **Linear driving and global quenching in homogeneous ion chain** | $d\propto$<br>$\begin{cases}\tau_Q^{-\frac{1}{4}};\quad if\ \frac{\delta_0}{\tau_Q}\ll\eta^3\\ \tau_Q^{-\frac{1}{3}};\quad if\ \frac{\delta_0}{\tau_Q}\gg\eta^3\\ unknown.\quad others\end{cases}$ | $d=$<br>$\left(\frac{3}{2}\tau_1^{\frac{1}{4}}\xi_1^{-1}\right)\tau_Q^{-\frac{1}{4}};\quad if\ \frac{\delta_0}{\tau_Q}\ll\eta^3$<br>$\tau_1^{\frac{1}{3}}\xi_1^{-1}\tau_Q^{-\frac{1}{3}};\quad if\ \frac{\delta_0}{\tau_Q}\gg\eta^3\ and\ \tau_Q\gg\beta$<br>*super-saturation region;* $\quad if\ \tau_Q\gg\beta\ and\ \beta\gg\frac{\delta_0}{\eta^3}$<br>*saturation region;* $\quad if\ \tau_Q\ll\beta$<br>*transition region.* $\quad$ *transition region.* |

cases, which is showed in *Table* 1(b). In fact, the extended formulation can deduce all the conclusions that KZM predicts, but it can give more information about defects varying than KZM. 1). KZM predicts in the overdamped regime $d\sim\tau_Q^{-1/3}$, but we show, this conclusion will not hold if $\eta^4\gg\delta_0^{3/2}\xi_1^3/(a_0^3\tau_1)$. Especially for the ion trap, if $\eta\gg 0.52\omega_0(\kappa_N/c_N)^{1/3}$, where $\kappa_N=\xi_N/\xi$, then the large range for the linear relation about $d\sim\tau_Q^{-1/3}$ will not appear. It is worth mentioning that for the continuous medium, $a_0\to 0,\ \beta$ in Eq. 4 equals zero. It means this inequation will not hold whenever, and we will observe the $d\sim\tau_Q^{-1/3}$ in the overdamped regime; 2). For the finite scale of continuous-transition-phase system, the defects producing will arrive the saturation region when quenching is very fast, which did not mention in KZM but can be observed in the both simulation and experiment for ion trap. The defects of a finite scaler of continuous-transition-phase system cannot be increase all the time when quenching rate increase because it has the maximal value that we can observe; 3), The supersaturation region, which is predicted from extended formulation for the finite scaler system, but have not be proved because kinks will annihilate and the exact statistics for them is difficult in ion trap when quenching is very fast. The mechanism for it is that the damping can prevent the disorder breaking. The quenching rate is bigger, the effect that particles are against disorder breaking is more obvious. The high damping will make the system enter ahead into the saturation region and then the supersaturation region forms.

**II). For linear driving rate and global quenching in harmonic ion trap.**

For inhomogeneous ion chain, the critical frequency change with the inter-particle distance $a_i$[25-27]. Different parts of ion chain will have different entrance time for critical point under the linear quenching. The propagation velocity in a different part is dependent on both time and position, $v = \xi_N(x,t)/\tau_N(x,t) = v(x,t)$. Therefore, Eq. (5) will transform into a integral equation. Because the critical frequency is dependent on the $a_i$, and every ion can be the nucleate site, therefore, we choose the average value, $\langle \hat{\xi}_{a_\mathrm{i}} \rangle$, to weigh the density of defects.

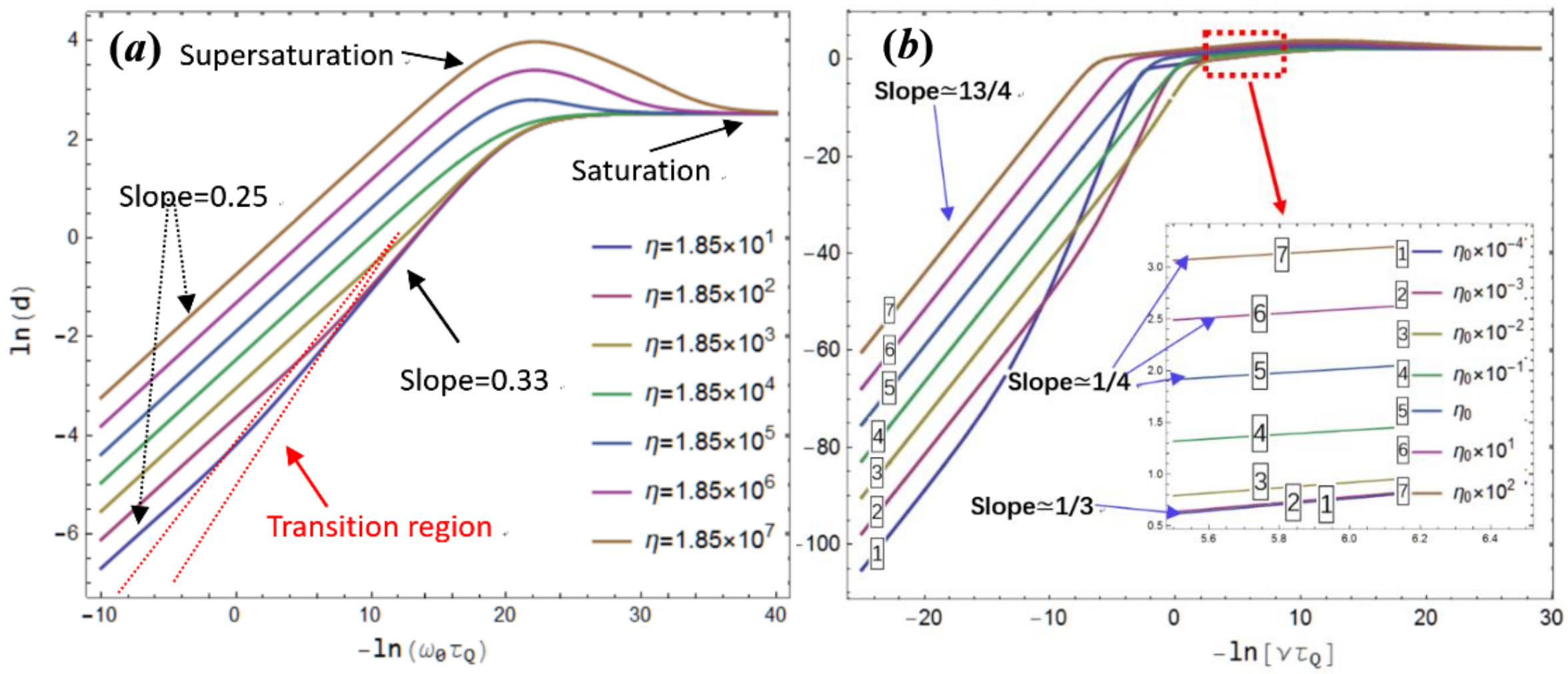

**Figure 2. The defects varying with the large range of $\tau_Q$ in homogeneous ion chain**, *graph.* a, and inhomogeneous ion chain, *graph* b. According these two graphics, we can see there are saturation region in both conditions. In the *graph* a, the supersaturation will appear as the damping coefficient $\eta$ increasing. In the *graph* b, we can see the slope changing with the damping coefficient. The asymptotic slope is 3.5 when $\tau_Q \rightarrow \infty$.

According to the result of Eq. (1), which show in the *fig.*1(a1) and *fig.*2(b), we can see the different results from KZM. 1). The slopes of 8/3, 4/3, 1/3 are not the tendency value in different conditions. They just the normal tangents in a smooth curve; 2). Under very slow quenching, the tendency value of slope is near 13/4; 3). Unlike the homogeneous ion chain, in which the defects are decrease as the damping increasing, the ion chain in harmonic trap will produce more defects as the damping increasing when damping exceeds a threshold, which can be seen in *fig.*2(b).

**III). For oscillation quenching in ion chain.**

Can we control the rate of defects producing? The answer is yes. One method is to manipulate the quenching rate linearly like KZM mentioned. It is the way to control the rate of defects production, but it is not effectively. When we want to greatly suppress the defects producing, we should keep the quenching in very low rate, which will make the evolution time very appreciable. If we want to control the rate of defects producing effectively, how should we do?

One way is the oscillation quenching. The oscillation quenching can control the rate of defects producing in given time. The effective control way for defects might be a board topic. Different oscillation way will lead to different efficiency in different condition. Here, we just give two examples, the $\sin^2(x)/x$ oscillation quenching in homogeneous ion chain and the $\sin(x^2)/x$ in harmonic ion trap.

For homogeneous ion chain, the formula of oscillation quenching is set as $\delta(t) = (-\lambda \sin^2 \omega t)/t$. For simplification, the overdamped condition and radial frequency driving just be considered here, namely, $\eta \gg 0.724\lambda\omega$. Under this quenching, the $\hat{\xi}$ is

$$\hat{\xi} = \frac{\xi_1 \lambda^{1/2} \sqrt{2\pi} F_{rs}\left(\frac{2}{\pi}\arcsin\left(\sqrt{c_N \eta/\lambda}\right)\right)}{c_N \eta \sqrt{\omega}}, \tag{7}$$

where the function $\mathrm{F_{rs}}(x)$ is the Fresnel integral. According to Eq. (7), we can see if $\lambda < c_N\eta$, the right side of Eq. (7) does not exist in real number set. That means there is no defects producing in this condition. To suppress the defects, we can set $\lambda < c_N\eta$ initially, and stop this oscillation quenching when $\xi(t)$ approaches the length of the ion chain $L$, and then quickly pull-down the radial frequency linearly. The time spent in oscillation process is $t_o = \tilde{n}\pi/\omega$, where $\tilde{n} \sim L(c_N \eta \sqrt{\omega})/2\xi_1 \lambda^{1/2}$, depending the length $L$ of ion chain. Usually, about few periodic oscillations can complete this process. Then the ions chain will form the zigzag morphology obviously with very low rate of defects producing. To complete this process, high oscillation frequency $\omega$ is necessary, but should not exceed the value $\eta/\lambda$.

Why this oscillation quenching can effectively suppress the rate of defects producing? As the thought of extended formulation, one nucleate site in ion chain is effective for ideal structural transition, but more than one will break the ideal transition and form the kinks. The oscillation quenching can motivate the effective nucleate site producing and suppress invalid nucleates. When the ion chain closes the critical point, ions of it have long survival time but very low propagate velocity $v_p$. It is contrary when the ions are far away the critical point. Therefore, we quickly pull-down the radial frequency, letting the ion chain far away the critical point to form the effective nucleate sit and quickly propagate this impact, then, pull back the radial frequency, letting the ions near the critical point to suppress the other probable nucleate sits producing. So repeatedly, the impact of until the effective nucleate site will increase and increase, finally the ideal zigzag morphology forming. Per the simulating with MD, in one thousand quenching loops with the same quenching time, the kinks under the linear driving quenching are 472 in the homogeneous ion chain with 22 ions.

However, under the oscillation quenching, the kinks rapidly decrease to 12. What is means? It means if we want to achieve this low defects in the same condition with linear quenching, we should increase the quenching time to $(472/12)^4 \sim 2.4\times10^6$ times.

For harmonic ion trap, this oscillation quenching is not appropriate because the critical frequency for inhomogeneous ion chain is not fix and the amplitude of the quenching $(-\lambda \sin^2 \omega t)/t$ decrease in every cycle. Some ions cannot cross the critical point all the time. Therefore, we choose a new oscillation quenching

$$\delta(x,t) = \omega_t^2 - \omega_c^2(x) = \frac{\delta_0}{\tau_Q}\left(\frac{t^2 + \lambda \sin \omega t^2}{t} - t_f\right). \tag{8}$$

As we can see, as the time increase, this oscillation quenching will be near the liear quenching. When $t \to 0$, the oscillation is very massive and the middle ion in ion chain will firstly cross the frozen area, forming the nucleate sit. When $t$ increase, the decreased amplitude will suppress the other nucleate sits producing. This oscillation can well suppress the defects. we show the effect of this oscillation in *fig.*3. Under this kind oscillation, the defects can be decreased into 0.1 or less.

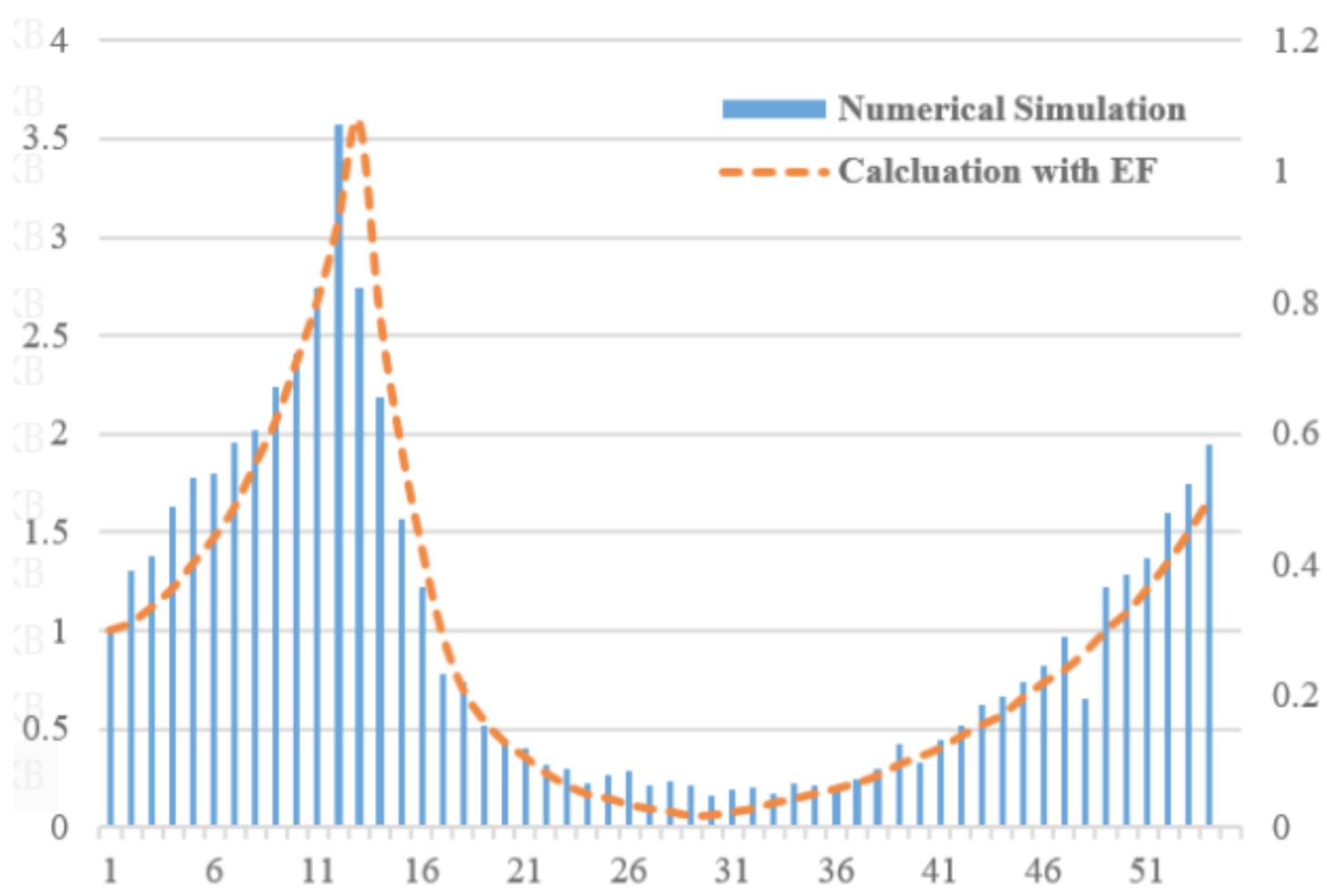

**Figure 3. the defects suppressed in harmonic ion trap under an oscillation quenching.** As the predict of extended formulation, up to 92.1% reduction in defects can be observable. In numerical simulation, we observe the 83.9% reduction.

## Conclusion and Acknowledge

Kibble-Zurek mechanism (KZM) is the theory underlying topological defect production in both classical and quantum phase transition. It makes greatly progress both in the condensed matter and quantum domain. However, it still has problems. We should use different formula in different conditions and we can only get the variation tendency of defects with quenching rate rather the exact results. The predicts of it are not in good agreement with the experimental data in harmonic ion trap, and some works in the quantum domain show the deviation from the KZ scaling law is found[28,29]. In this paper, we pointed out its drawback and proposed a new formulation to overcome the problems encountered.

We show the possible unified form, extended formulation, can be compatible with KZ scaling law in homogeneous structure under linear and global quenching (for the quenching rate with the form $\epsilon = \epsilon_0 |t| \mathrm{sign}(t)$, the two are also compatible with each other). For other systems in other conditions, the extended formulation provides more exact descriptions. In this article, we compare the extended formulation with KZM in ion trap system and show the obvious advantage of the extended formulation. Furthermore, to exhibit its great potential applications in very fast quantum state manipulations, we take two kinds oscillation quenching as examples. Under these oscillations, we observe the obvious defects-suppress and that is in good agreement with the predicts of the extended formulation.

We expound the extended formulation mainly in the ion trap system, but its paradigm is general and it can be hopefully used in other condensed matter systems or quantum models. We very appreciate Prof. Ming Gong for helpful discussions and this work is supported by National Science Foundation 11547165 .